\documentclass[aps,reprint,superscriptaddress,nofootinbib]{revtex4-1}

\usepackage{graphicx}
\usepackage{dcolumn}
\usepackage{bm}
\usepackage{amssymb}
\usepackage{amsmath}

\begin{document}


\title{Coarsening dynamics of binary liquids with active rotation}

\author{Syeda Sabrina}
\affiliation{Department of Chemical Engineering, Pennsylvania State University, University Park, PA 16802, USA.}

\author{Matthew Spellings}
\affiliation{Department of Chemical Engineering, University of Michigan, Ann Arbor, MI 48109, USA.} 
\affiliation{Biointerfaces Institute, University of Michigan, Ann Arbor, MI 48109, USA.}

\author{Sharon C. Glotzer}
\email{sglotzer@umich.edu}
\affiliation{Department of Chemical Engineering, University of Michigan, Ann Arbor, MI 48109, USA.} 
\affiliation{Biointerfaces Institute, University of Michigan, Ann Arbor, MI 48109, USA.} 
\affiliation{Department of Material Science and Engineering, University of Michigan, Ann Arbor, MI 48109, USA.}

\author{Kyle J. M. Bishop}
\email{kjmbishop@engr.psu.edu}
\affiliation{Department of Chemical Engineering, Pennsylvania State University, University Park, PA 16802, USA.}




\date{\today}

\begin{abstract}
Active matter comprised of many self-driven units can exhibit emergent collective behaviors such as pattern formation and phase separation in both biological (\textit{e.g.}, mussel beds) and synthetic (\textit{e.g.}, colloidal swimmers) systems. 
While these behaviors are increasingly well understood for ensembles of linearly self-propelled ``particles'', less is known about the collective behaviors of active rotating particles where energy input at the particle level gives rise to rotational particle motion.
A recent simulation study [Nguyen \textit{et al.},  \textit{Phys. Rev. Lett.}, 2014, \textbf{112}, 075701] revealed that active rotation can induce phase separation in mixtures of counter-rotating particles in 2D.
In contrast to that of linearly self-propelled particles, the phase separation of counter-rotating fluids is accompanied by steady convective flows that originate at the fluid-fluid interface.
Here, we investigate the influence of these flows on the coarsening dynamics of actively rotating binary liquids using a phenomenological, hydrodynamic model that combines a Cahn-Hilliard equation for the fluid composition with a Navier-Stokes equation for the fluid velocity.
The effect of active rotation is introduced though an additional force within the Navier-Stokes equations that arises due to gradients in the concentrations of clockwise and counter-clockwise rotating particles.
Depending on the strength of active rotation and that of frictional interactions with the stationary surroundings, we observe and explain new dynamical behaviors such as ``active coarsening'' \textit{via} self-generated flows as well as the emergence of self-propelled ``vortex doublets''. 
We confirm that many of the qualitative behaviors identified by the continuum model can also be found in discrete, particle-based simulations of actively rotating liquids. 
Our results highlight further opportunities for achieving complex dissipative structures in active materials subject to distributed actuation.

\end{abstract}

\pacs{Valid PACS appear here}
\maketitle


\section{\label{sec:level1}Introduction}
The distributed conversion of energy into motion within ensembles of many self-propelled units can lead to complex collective behaviors operating outside the constraints of thermodynamic equilibrium \cite{Marchetti2013}. Well-studied examples of such active matter include migrating organisms \cite{Vicsek2012}, the cell cytoskeleton \cite{Joanny2009,Sumino2012}, driven granular materials \cite{Sapozhnikov2003,Narayan2007}, and self-phoretic colloids \cite{Howse2007,Ibele2009,Kagan2011,Theurkauff2012,Palacci2013,Buttinoni2013}.
In many of these systems, the activity of the individual units can lead to phase separation and coexistence even in the absence of attractive interactions.
This behavior is clearly illustrated by simple physical models such as that of active Brownian particles (ABPs), in which hard spheres move at a constant speed in a direction subject to rotational Brownian motion \cite{Howse2007}.
ABPs are known to phase separate in 2D \cite{Fily2012,Redner2013,Stenhammar2013} and 3D \cite{Stenhammar2014,Wysocki2014} due to a kinetic trapping mechanism, whereby particles incident on the surface of a condensed phase are “trapped” by other incoming particles \cite{Redner2013}. 
More generally, activity-induced phase separation of self-propelled “particles” is expected whenever the average particle velocity decreases sufficiently rapidly with particle density \cite{Tailleur2008,Cates2013}.
This basic mechanism is believed to underlie phase separation in such disparate systems as mussel beds \cite{Liu2013}, bacterial colonies \cite{Cates2010}, and active colloids \cite{Buttinoni2013}.

Activity-induced phase separation has also been observed in systems of rotating  particles, in which otherwise identical, gear-like disks are driven to rotate in opposite directions \cite{Nguyen2014}.
Active rotation induces effective interactions between the particles that can cause their segregation into counter-rotating, fluid and crystalline domains \cite{Nguyen2014}. 
Similar behaviors have been observed in simulations of spherical particles rotating within a fluid, where viscous shear forces couple the rotational motions of neighboring particles \cite{Goto2015,Yeo2014,Yeo2015}.
Experimentally, there exists several promising mechanisms by which to rotate colloidal components using magnetic fields \cite{Grzybowski2000,Tierno2008}, electrokinetic flows \cite{Boymelgreen2014}, self-phoretic motions \cite{Qin2007,Wang2009,Ebbens2010}, or circularly polarized light \cite{Cheng2002}. 
Recent studies have shown that single component systems of magnetically rotated colloids \cite{Yan2015} or self-rotating bacteria \cite{Petroff2015} can segregate into high and low density phases driven primarily by attractive dipolar or hydrodynamic interactions, respectively.

Importantly, the unmixing of actively rotating particles is accompanied by steady convective flows that originate along the interface separating the counter-rotating domains \cite{Nguyen2014}.
Under appropriate conditions, these activity-driven flows are expected to influence the coarsening of actively rotating fluids and may lead to new types of dynamically-organized structures.
By contrast, the coarsening of linearly self-propelled particles such as ABPs exhibit strong similarities to that of passive liquids \cite{Wittkowski2014}, in which self-similar domains of size $R$ grow in time as $R\propto t^{1/3}$. 
Thus, while different forms of microscopic activity can lead to similar mesoscale behaviors (e.g., phase separation), these differences may contribute to qualitatively different dynamical behaviors at the macroscale (e.g., coarsening dynamics).

\begin{figure}[h]

  \includegraphics{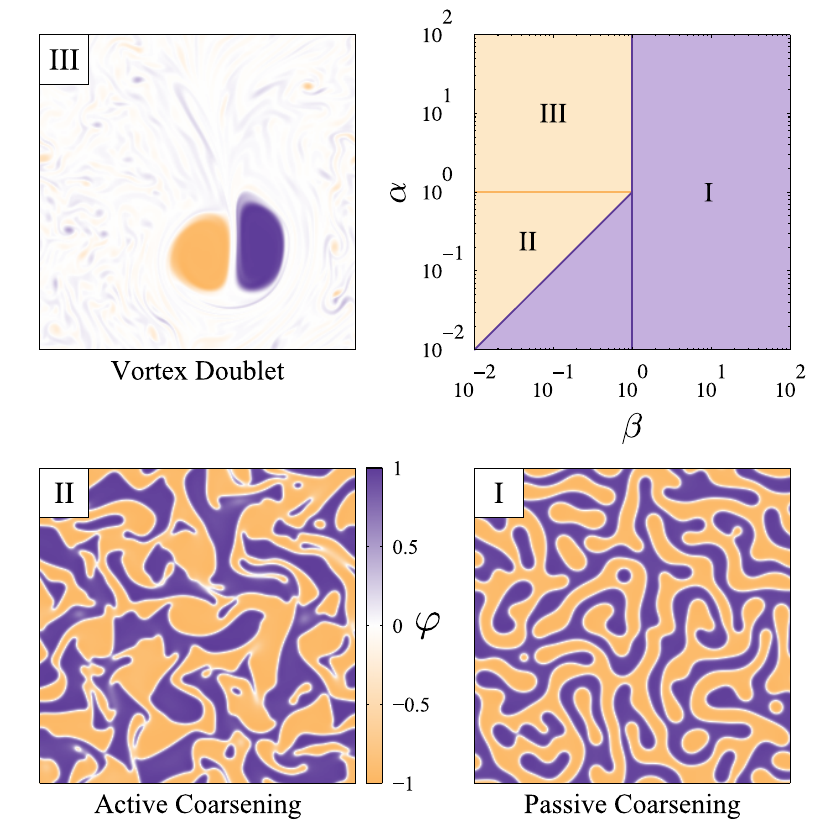}
  \caption{Phase diagram in the $\alpha\beta$-plane illustrating the different dynamical regimes (top right). Here, $\alpha$ measures the strength of active rotation, while $\beta$ measures that of frictional damping against the stationary surroundings (see text for details). One representative snapshot of the compositional order parameter $\varphi$ is shown for each regime. }
  \label{fig:Fig1}
\end{figure}

Here, we investigate the dynamics of liquid-liquid unmixing in a binary fluid subject to active rotation using a phenomenological, phase-field model based on the convective Cahn-Hilliard equation \cite{Jasnow1996,Anderson1998}.
In this description, the rotational actuation of the two components in opposite directions introduces additional forces within the Navier-Stokes equations governing fluid motion.
Depending on the strength of active rotation and that of frictional interactions with the stationary surroundings, we observe three distinct dynamical regimes as summarized in Figure \ref{fig:Fig1}.  
For strong frictional damping, coarsening of the counter-rotating domains is identical to that of a passive fluid without active rotation (in agreement with previous particle-based simulations \cite{Nguyen2014}).
By contrast, when frictional forces are relaxed, the system exhibits new dynamical behaviors such as ``active coarsening'' driven by convective flows induced by the rotation of the particles as well as the emergence of self-propelled ``vortex doublets''.
We use numerical simulation along with scaling arguments to characterize the system's dynamics within each flow regime.
Additionally, we show that many of the qualitative behaviors identified by the continuum model can also be found in discrete, particle-based simulations of actively rotating liquids.
These results highlight opportunities for achieving complex dissipative structures by directing collective excitations within active matter.

\section{\label{sec:level2}Model Dynamics}
The coarsening dynamics of binary liquids with active rotation is studied using two models: a continuum, hydrodynamic model and a microscopic, particle-based model. 
The former is an extension of previous phase field models \cite{Jasnow1996,Anderson1998} for two phase flow that accounts for the active rotation of the fluid components.  
The microscopic model \cite{Nguyen2014,Spellings2015} describes the Langevin dynamics of hard, gear-like particles, which are driven to rotate in opposite directions.
Here, we limit our investigation to two-dimensional systems; however, both models can be readily extended to three-dimensions.

\subsection{\label{sec:leve21}Continuum Model}
In the continuum approach, we consider a binary fluid in which the local composition is characterized by an order parameter $\varphi$ governed by the convective Cahn-Hilliard equation \cite{Jasnow1996,Anderson1998},
\begin{equation}
\frac{\partial \varphi }{\partial t}+\nabla \cdot (\varphi \mathbf{v})=M{{\nabla }^{2}}\mu,\label{eq:CH}
\end{equation}
where $\mathbf{v}$ is the fluid velocity, $M$ is a mobility coefficient, and $\mu$ is the chemical potential. 
Physically, the composition $\varphi(x,y,t)$ describes the relative amount of counter-clockwise-rotating components ($\varphi>0$) and clockwise-rotating components ($\varphi<0$) at a given point in space and time. 
For simplicity, we assume the chemical potential is of the form
\begin{equation}
    \mu =-r\varphi +\lambda {{\varphi }^{3}}-K{{\nabla }^{2}}\varphi,\label{eq:CP} 
\end{equation}
where $r$, $\lambda$, and $K$ are positive coefficients. 
These coefficients determine the thickness $(K/r)^{1/2}$ of the interface separating two phases of composition $\varphi=\pm(r/\lambda)^{1/2}$. 
We emphasize that this simple model does not attempt to explain the \textit{origins} of phase separation driven by active rotation.
Instead, we assume phase separation \textit{a priori} and focus on the role of activity on the dynamics with which these rotating phases coarsen in time. 

To describe the activity-driven flows, we further assume that the fluid is incompressible, Newtonian, and ``symmetric'' such that the bulk properties of the two phases are equal -- in particular, the density $\rho$ and viscosity $\eta$.  
Under these conditions, conservation of mass and momentum imply that 
\begin{gather}
    \nabla\cdot\mathbf{v} = 0,\label{eq:Continuity}\\
    \rho\frac{d\mathbf{v}}{dt} = -\nabla p + \eta \nabla^2 \mathbf{v} + \mu \nabla \varphi + \nabla\times(\varphi\boldsymbol\tau) - b\mathbf{v}. \label{eq:NS}
\end{gather}
In addition to the usual pressure and viscous forces present in the Navier-Stokes equation, equation (\ref{eq:NS}) incorporates forces due to (i) capillarity \cite{Jasnow1996}, (ii) active rotation, and (iii) frictional drag, respectively.
In particular, we consider that the two components of the fluid are driven to rotate in opposite directions by a torque density $\varphi\boldsymbol\tau$, which is proportional to the order parameter $\varphi$ and to a constant vector $\boldsymbol\tau$ that describes the magnitude and direction of rotation. 
These local torques combined with spatial variations in the composition give rise to forces that act parallel to the interface separating the counter-rotating phases \cite{Rosensweig2004}.
In our 2D simulations, the fluid moves in the $xy$-plane with active rotation in the $z$-direction ($\boldsymbol\tau=\tau\mathbf{e}_z$).
Physically, the system can be thought to represent an ensemble of active particles moving and rotating above a planar substrate as is often the case in experimental realizations of active matter in 2D. 
To account for interactions between the particles and the underlying substrate, we include a frictional force in equation (\ref{eq:NS}) characterized by a constant friction coefficient $b$ .

At this point, it is convenient to non-dimensionalize the governing equations using characteristic scales for the interfacial thickness $(K/r)^{1/2}$, the time of unmixing $K/Mr^2$, the equilibrium composition $(r/\lambda)^{1/2}$, and the chemical potential $(r^3/\lambda)^{1/2}$. 
In dimensionless units, equations (\ref{eq:CH}) and (\ref{eq:NS}) reduce to
\begin{gather}
\frac{\partial \varphi }{\partial t}+\mathbf{v}\cdot \nabla \varphi = \nabla^2 \left( -\varphi + \varphi^3 - \nabla^2 \varphi \right), \label{eq:CH_nounits}
\\
Re \frac{d\mathbf{v}}{dt} = -\nabla p + \nabla^2 \mathbf{v} + Ca^{-1}\mu \nabla \varphi +\alpha \nabla \times (\varphi \mathbf{e}_z) - \beta \mathbf{v},\label{eq:NS_nounits}
\end{gather}
where $Re=\rho M r/\eta$ is a is a Reynolds number, $Ca=M\lambda\eta/K$ is a capillary number, and the dimensionless coefficients $\alpha$ and $\beta$ characterize the strength of active rotation and frictional drag, respectively. 
In this paper, we focus exclusively on the low Reynolds number limit ($Re\rightarrow0$) and neglect capillary forces ($Ca^{-1}\rightarrow0$) such that fluid flow is driven solely by the active rotation of the fluid components.    

The governing equations (\ref{eq:CH_nounits}) and (\ref{eq:NS_nounits}) are solved numerically on a square domain ($L\times L$) with periodic boundaries using a semi-implicit Fourier spectral method \cite{Zhu1999} for different values of the parameters $\alpha$ and $\beta$.
Initially, the system is prepared in a homogeneous state, in which the composition at each point is assigned a random value drawn uniformly from the interval [$-0.1$, $0.1$]. 
Depending on the strength of active rotation $\alpha$ and frictional drag $\beta$, this model exhibits a variety of different coarsening mechanisms ranging from passive, diffusive coarsening to active coarsening and the emergence of ``vortex doublets'' (Fig. \ref{fig:Fig1}).

\subsection{\label{sec:level22}Microscopic Model}

To confirm the generality of active coarsening in rotating fluids, we study an analogous particle-based system whereby collections of hard, gear-shaped ``spinners'' are driven to rotate in opposite directions by an applied torque \cite{Nguyen2014, Spellings2015}.
Each spinner contains five circular disks of radius $\sigma$ fixed symmetrically about a central disk of radius $3\sigma$.
The dynamics of these composite particles is governed by the following Langevin equation for the velocity of the $i$th disk
\begin{equation}
    m \frac{d\mathbf{v}_i}{d t} = \mathbf{F}_i -\gamma\mathbf{v}_i +\mathbf{F}_i^R,
    \label{eq:LE}
\end{equation}
where $m$ is the mass of each disk, $\mathbf{F}_i$ and $\mathbf{F}^R_i$ represent deterministic and stochastic forces, and $\gamma$ is a frictional drag coefficient.
The deterministic forces $\mathbf{F}_i$ contain both active and passive contributions.
First, all spinners are driven to rotate by a constant torque $\tau_i =\pm\tau$ with equal numbers rotating in each direction.
Additionally, spinners interact both through a repulsive contact potential and through a short ranged attraction between like-rotating spinners.
The latter is included to ensure phase separation even in the absence of active rotation by analogy to the continuum model, although it was not considered in previous works \cite{Nguyen2014}.
The stochastic force, $\mathbf{F}_i^R=\sqrt{2\gamma k_B T}\mathbf{X}(t)$, ensures that the system approaches thermal equilibrium at temperature $T$ in the absence of active rotation. 
Langevin dynamics simulations were performed on graphic processing units (GPUs) with the HOOMD-blue \cite{Anderson2008a,Nguyen2011a} software package for $16,384$ spinners in the system.

Although a rigorous connection between the microscopic and continuum models is lacking and outside the scope of this paper, we use order-of-magnitude reasoning to identify dimensionless parameters $\alpha'$ and $\beta'$ in the microscopic model that are analogous to $\alpha$ and $\beta$ in the continuum model.
Specifically, $\alpha'=\tau/k_B T$ measures the strength of active rotation relative to the thermal energy, whereas $\beta'=\gamma\sigma/\sqrt{m k_B T}$ measures the strength of frictional damping.
Below, all results from the microscopic model are presented in dimensionless form using characteristic scales $\sigma$, $\sigma (m/k_B T)^{1/2}$, and $k_B T$ for length, time, and energy, respectively.

\section{\label{sec:level3}Results and Discussion}

We first use the continuum model to map out three qualitatively distinct parameter regimes as summarized in Figure \ref{fig:Fig1}. 
We discuss each regime in turn and provide detailed scaling arguments to explain the behaviors observed in the simulations.
Building on insights from the continuum model, we reproduce many -- though not all -- of the qualitative coarsening behaviors using the microscopic model.

\subsection{\label{sec:level31}Strong Damping ($\beta\gg1$)}
\begin{figure}[h]

  \includegraphics{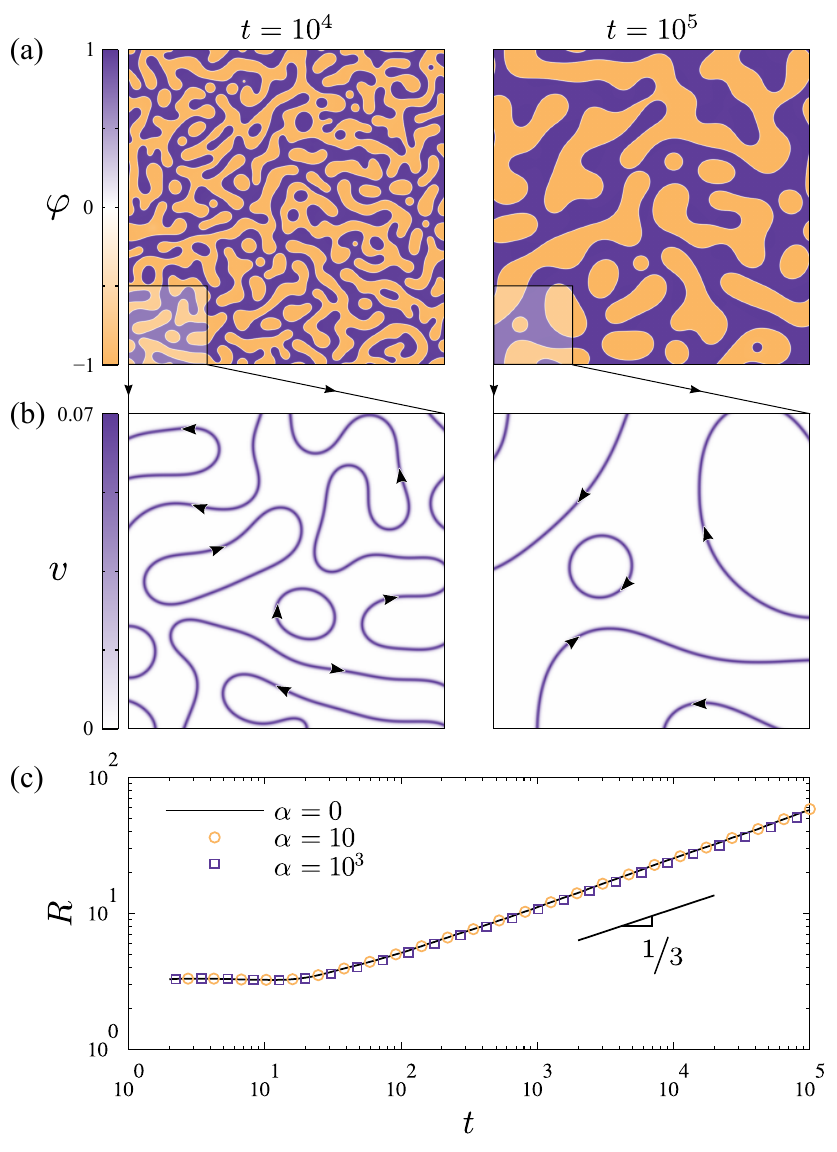}
  \caption{\textbf{Passive coarsening, $\beta\gg1$}. (a) Compositional order parameter $\varphi(x,y)$ at times $t = 10^4$ and $10^5$ for parameters $\alpha = 10$ and $\beta = 10^2$; the size of simulation cell is $L = 1024$. (b) Velocity field $v(x,y)$ corresponding to the insets in (a). Arrows show the direction of fluid flow. (c) Domain size $R$ as a function of time for $\beta = 10^2$ and $\alpha = 0$, $10$, and $10^3$; here, $R$ is defined as the first zero of the radial pair correlation function, $g(R) = 0$ \cite{Zhu1999}.}
  \label{fig:Fig2}
\end{figure}

In the presence of strong frictional damping ($\beta\gg1$), the coarsening dynamics of the active fluid is independent of the strength of active rotation (i.e., of $\alpha$) and identical to that of a passive fluid, for which $\alpha=0$ (Fig. \ref{fig:Fig2}). 
This ``passive coarsening'' regime has been studied extensively using the Cahn-Hilliard equation in the absence of fluid flow \cite{Bray2002}. 
At short times ($t\ll1$), the initially homogeneous fluid undergoes an instability characterized by a wavenumber $k=2^{-1/2}$, which grows in time at a rate $1/4$ until the formation of bulk domains with composition $\varphi\approx\pm1$ separated by an interfacial region of unit thickness. 
At longer times ($t\gg1$), these domains grow in size as $R\sim t^{1/3}$ due to small composition gradients ($\Delta\varphi\sim R^{-1}$) which drive diffusive fluxes ($j\sim\Delta\varphi/R$) that act to grow the domains ($dR/dt\sim j$) and reduce the curvature of the interface (Fig. \ref{fig:Fig2}a). 
Here, the domain size $R$ is defined as the first root of the radial pair correlation function \cite{Zhu1999} unless otherwise stated.

In this regime, the active rotation of the fluid drives convective flows along the the interface separating the counter-rotating domains (Fig. \ref{fig:Fig2}b). 
The dominant terms of equation (\ref{eq:NS_nounits}) are $\alpha\nabla\times(\varphi\mathbf{e}_z) \approx \beta\mathbf{v}$, such that forces due to active rotation are everywhere balanced by frictional drag.
As a result, all flows are directed perpendicular to gradients in the order parameter, and the effects of convective transport are negligible (i.e., $\mathbf{v}\cdot\nabla\varphi\approx0$ in equation (\ref{eq:CH_nounits})). 
Consequently, the domain size $R$ increases as $R \sim t^{1/3}$ independent of both $\alpha$ and $\beta$ in quantitative agreement with passive diffusive coarsening (Fig. \ref{fig:Fig2}c).

\subsection{\label{sec:level32}Weak Damping \& Weak Rotation ($\beta\ll1$, $\alpha\ll1$)} 
\begin{figure}[h]
\includegraphics{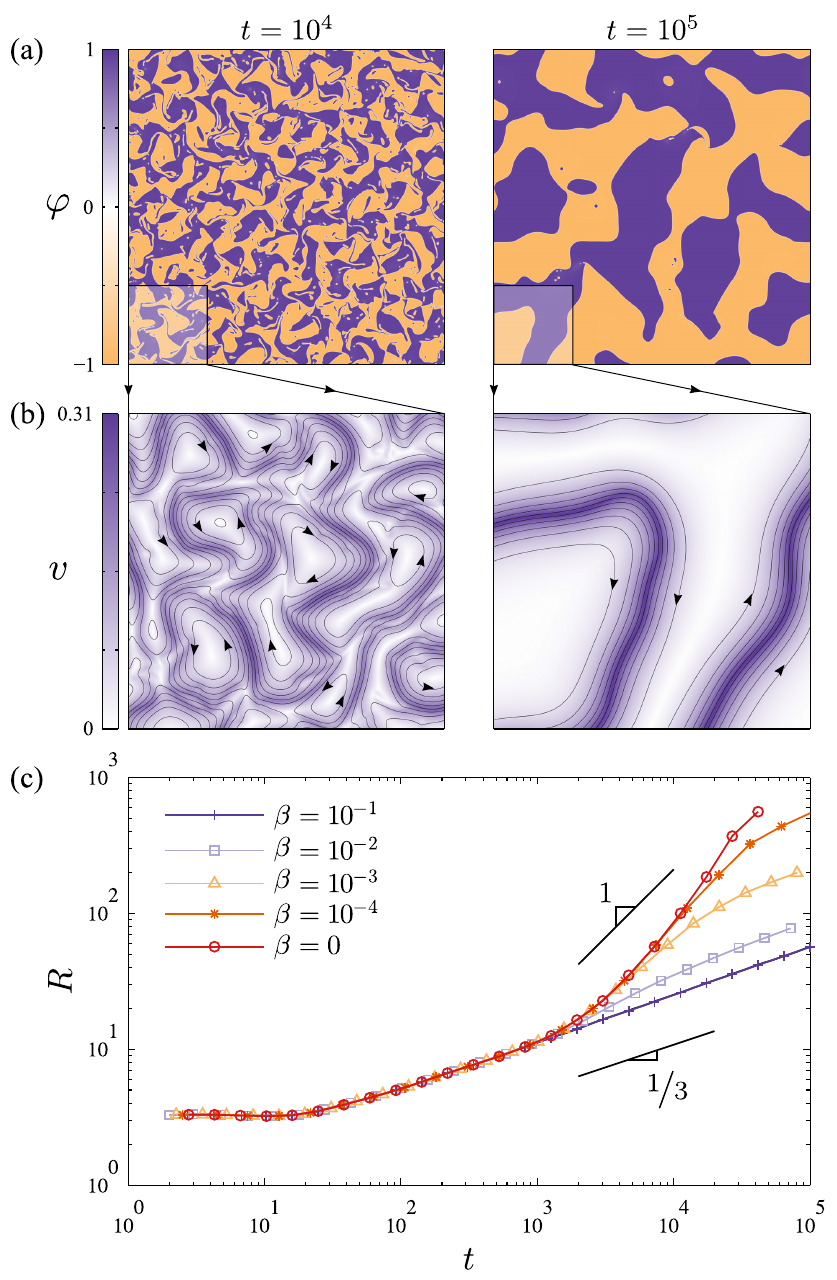}
\caption{\textbf{Active coarsening, $\beta\ll1$ and $\alpha\ll1$}. (a) Order parameter $\varphi(x,y)$ at times $t = 10^4$ and $10^5$ for parameters $\alpha = 10^{-2}$ and $\beta = 10^{-3}$; the size of simulation cell is $L = 2048$. (b) Velocity field $v(x,y)$ corresponding to the insets in (a). Arrowheads show the direction of fluid flow. (c) Domain size $R$ as a function of time for $\alpha=10^{-2}$ and different amounts of frictional damping $\beta$.}
\label{fig:Fig3}
\end{figure}

For weak frictional damping ($\beta\ll1$), flows due to active rotation are no longer confined to the interface but rather extend into the bulk domains to influence the dynamics of unmixing (Fig. \ref{fig:Fig3}). 
The morphology of the growing domains (Fig. \ref{fig:Fig3}a) is visibly different from that due to passive coarsening: arrays of counter-rotating vortices (Fig. \ref{fig:Fig3}b) create thin filaments that break-up and merge with the larger domains.
The characteristic domain size $R$ increases faster with time than expected by diffusive coarsening alone (Fig. \ref{fig:Fig3}c).

To understand these qualitative observations in more detail, consider that rotation within the bulk creates an interfacial stress of order $\alpha$ in a direction tangent to the interface.
This active stress is balanced by the viscous stress $U/\ell$, where $U$ is a characteristic velocity, and $\ell$ is a length scale over which the velocity falls to zero. 
For small domains ($R\ll\beta^{-1/2}$), velocity gradients extend throughout the bulk such that $\ell\sim R$ and $U\sim\alpha R$.
As the domains grow larger ($R\gg\beta^{-1/2}$), the velocity decays exponentially with distance from the interface over a length $\beta^{-1/2}$ due to frictional drag; the velocity approaches a constant value $U\sim\alpha \beta^{-1/2}$ (Fig. \ref{fig:Fig3}b).

Using these estimates for the fluid velocity, we introduce a P\'eclet number, $Pe = \ell U$, which characterizes the relative importance of convective and diffusive transport on the coarsening of the domains.\footnote{In dimensional units, the P\'eclet number takes the more familiar form of $Pe = \ell U / M r$ where $Mr$ is identified as the diffusivity.}
For small P\'eclet number ($Pe\ll1$), activity-driven flows do not affect the coarsening dynamics, which is analogous to that of a passive fluid. 
Using the above estimates for the fluid velocity, this condition implies that small domains, $R\ll\alpha^{-1/2}$, are unaffected by active rotation.
By contrast, domains that grow larger than a critical size, $R^{*}\sim\alpha^{-1/2}$, induce flow velocities capable of influencing the coarsening dynamics. 
This effect is illustrated in Fig. \ref{fig:Fig3}c which shows that the domain size $R(t)$ follows that of the passive fluid for $R<10$ when $\alpha=10^{-2}$. 
Beyond the critical size $R^{*}$, coarsening accelerates due to convection driven by the rotating fluid. 
In this regime, domain growth is expected to scale as $dR/dt \sim U\Delta\varphi \sim \alpha$, where $\Delta\varphi\sim R^{-1}$ is the magnitude of curvature-induced variations in composition.

Eventually, however, the rate of coarsening slows as the domain size $R$ grows larger than the length $\ell\sim\beta^{-1/2}$, which characterizes the decay of velocity with distance from the interface. 
Under these conditions ($R\gg\beta^{-1/2}$), flows are increasingly confined within a thin interfacial region and no longer influence the rate-limiting process of diffusion throughout the bulk domain. 
As a result, the domain growth returns to the diffusive scaling, $R\propto t^{1/3}$, at long times (Fig. \ref{fig:Fig3}c). 

To summarize, ``active coarsening'' occurs when (i) the P\'eclet number is large, and (ii) activity-driven flows extend throughout the bulk domains. 
These conditions are satisfied provided that the domain size is in the range $\alpha^{-1/2}\ll R \ll \beta^{-1/2}$. 
This dynamical regime is denoted by region II of the phase diagram in Fig. \ref{fig:Fig1}.

\subsection{\label{sec:level33}Zero Damping \& Weak Rotation ($\beta\rightarrow0$, $\alpha\ll1$)} 

\begin{figure}[h]

\includegraphics{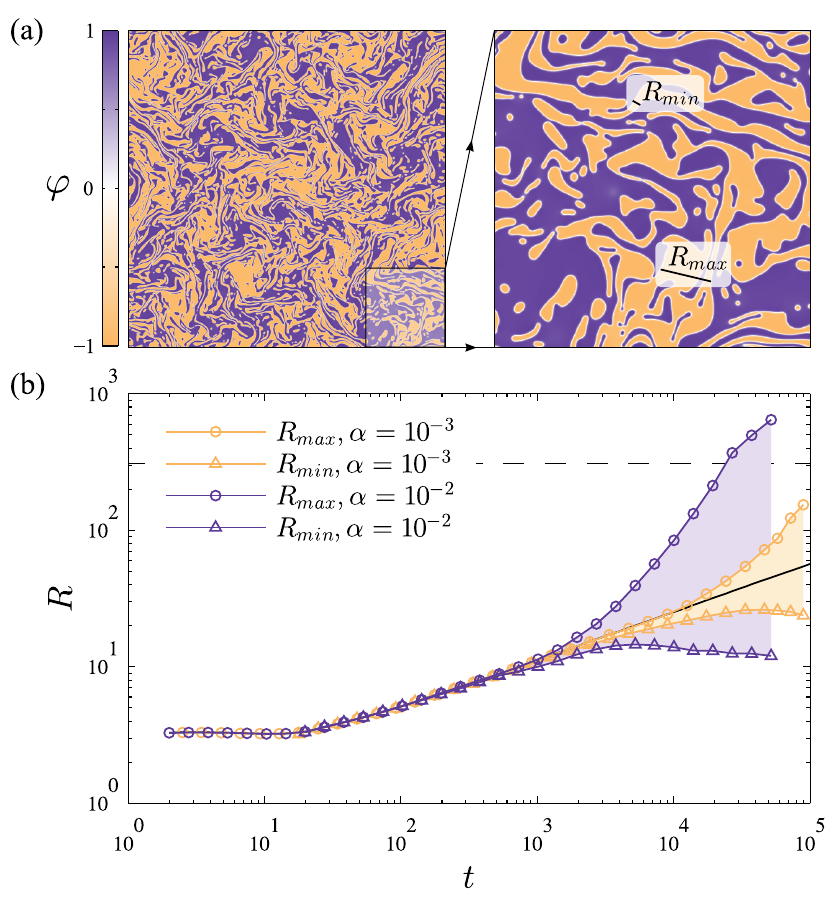}
\caption{\textbf{Active coarsening, $\beta\rightarrow 0$ and $\alpha\ll1$}. (a) Order parameter $\varphi(x,y)$ at time $t = 2\times10^4$ for $\alpha = 10^{-2}$ and no frictional damping, $\beta = 0$; the size of simulation cell is $L = 2048$. The right image shows a magnified view highlighting the two characteristic lengths, $R_{min}$ and $R_{max}$, described in the text. (b) Length scales, $R_{min}$ and $R_{max}$, as a function of time for $\alpha=10^{-2}$ and $\alpha=10^{-3}$ in the absence of frictional damping, $\beta=0$. The dashed black curve corresponds to $0.1L$ beyond which the finite size of the simulation domain becomes important; the solid black curve shows the domain size evolution for a passive fluid.}
  \label{fig:Fig4}
\end{figure}

To better understand the ``active coarsening'' regime, we examine the limit of zero frictional damping, $\beta\rightarrow0$, such that interfacial stresses due to active rotation are propagated by viscosity throughout the bulk domains -- regardless of their size (Fig. \ref{fig:Fig4}).
The domain structure is no longer characterized by a single length scale in contrast to the self-similar structures formed by ``passive coarsening''. 
Instead, we observe a spectrum of different length scales spanning a finite range from $R_{min}$ to $R_{max}$. 

The lower bound can be estimated as $R_{min} \sim A / C$, where $A$ is the total area, and $C$ is the length of interface separating the bulk domains. 
Physically, $R_{min}$ describes the width of the filamentous structures that are repeatedly drawn from the edges of the larger rotating domains.
Because these structures are shaped by convective flows, their size must be sufficiently large to achieve P\'eclet numbers of order unity -- that is, $Pe=\alpha R^2_{min}\sim1$ such that $R_{min}\sim\alpha^{-1/2}$. 
This scaling result is supported by numerical simulations (Fig. \ref{fig:Fig4}b), which reveal that $R_{min}$ remains roughly constant throughout the coarsening process.

The larger length scale $R_{max}$ is evaluated like $R$ above as the first root of the pair correlation function.
Physically, the composition at two points separated by distances less than $R_{max}$ are positively correlated; however, the strength of these correlations is considerably less than those observed for passive coarsening owning to heterogeneity within these larger domains. $R_{max}$ grows roughly linearly in time until \textit{ca.} $0.1L$, beyond which the finite size of the simulation cell begins to significantly influence the systems' dynamics.

In contrast to systems with frictional damping, which ultimately phase separate into bulk domains of arbitrary size, the multi-scale structures that arise in the zero-friction limit appear to avoid macroscopic phase separation indefinitely. 
Instead, active rotation continuously stretches and folds the growing domains in an effort to ``mix'' the fluid while it stubbornly attempts to ``unmix''. 
Competition between these two processes cause the formation of the smaller structures of order $R_{min}$, which appear to persist indefinitely (barring finite size effects).

\subsection{\label{sec:level34}Zero Damping \& Strong Rotation ($\beta\rightarrow0$, $\alpha\gg1$)}

\begin{figure}[h]

  \includegraphics{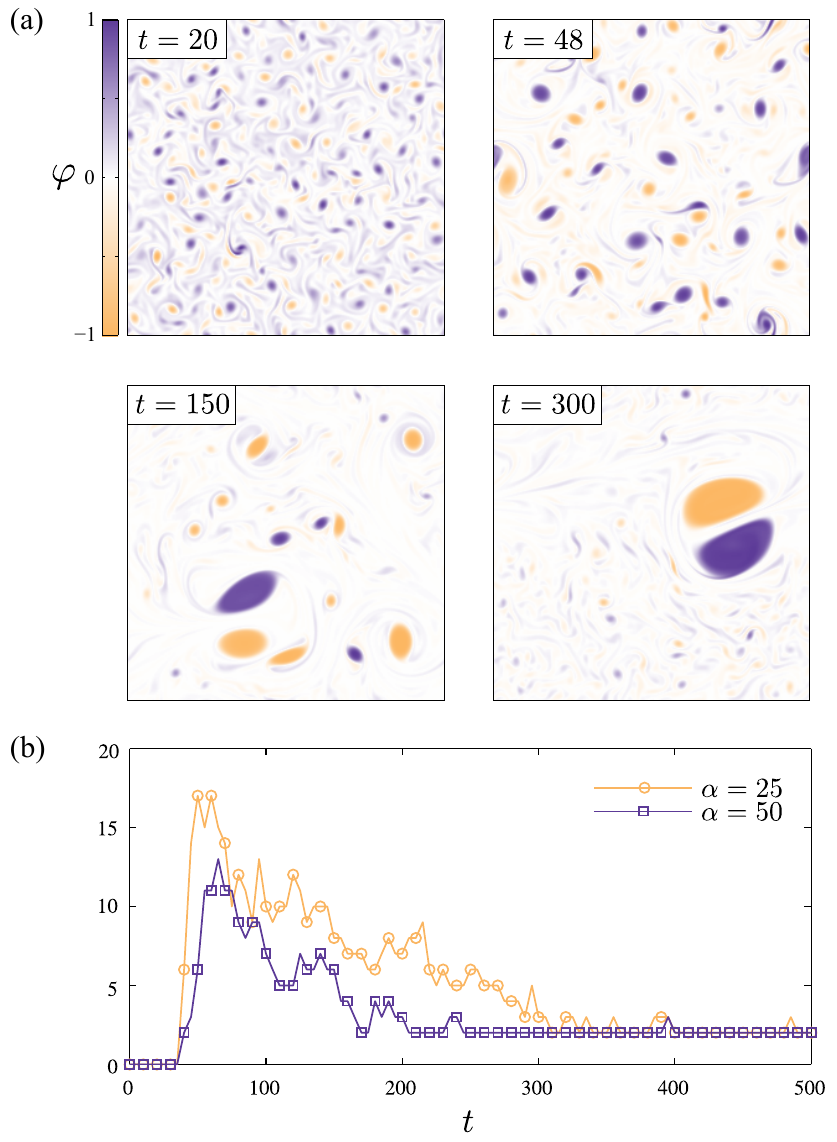}
  \caption{\textbf{Vortex doublet, $\beta\rightarrow 0$ and $\alpha\gg1$}. (a) Order parameter $\varphi(x,y)$ at times $t = 20$, 48, 150, and 300 for parameters $\alpha = 50$ and $\beta = 0$; the size of simulation cell is $L = 256$. Number of vortices as a function of time for $\beta = 0$ and $\alpha = 25$ and $50$. Here, a vortex is identified as a localized region in which the composition is $|\varphi| \geq 0.3$.}
  \label{fig:Fig5}
\end{figure}

As the magnitude of active rotation is increased such that $\alpha \gg 1$, the system transitions to a new dynamical regime characterized by the nucleation of localized vortices that move, interact, and combine within an otherwise homogeneous fluid (Fig. \ref{fig:Fig5}).
Ultimately, a single pair of counter-rotating vortices -- a ``vortex doublet'' -- emerges and propels itself autonomously throughout the domain, thereby mixing the fluid and preventing further phase separation.

In this regime, activity-driven flows begin to shape the dynamics of the composition \textit{prior} to the formation of the bulk phases. 
At these early times, the characteristic fluid velocity scales as $U\sim\alpha\Delta\varphi$, where $\Delta\varphi$ characterizes the magnitude of composition variations over a unit length (corresponding to the size $k^{-1}$ of the fastest growing mode). 
As above, convection begins to compete with diffusive transport when the P\'eclet number is of order unity, $Pe \sim \alpha\Delta\varphi \sim 1$. 
For strong rotation, activity-driven flows become significant even for partial phase separation -- that is, when $\Delta\varphi \sim \alpha^{-1} \ll1$. 

Importantly, these convective flows have the potential to inhibit the further unmixing of the two fluid components. In the absence of active rotation, fluid unmixing proceeds exponentially as $\Delta\varphi\propto\exp(t/4)$. 
To inhibit phase separation, the shear rate in the fluid must exceed the rate of unmixing. 
Partial phase separation results in activity-driven flows with shear rates of order $\alpha\Delta\varphi$.
Thus, when the extent of unmixing reaches a critical value -- namely, $\Delta\varphi > \alpha^{-1}$ -- the resulting flows will act to oppose further unmixing. 

Nevertheless, the spatial heterogeneity of the shear field allows for the nucleation of small vortices in locations with lower shear rates.
As vortices form, they create regions of low shear (but high vorticity) in their interior that allow for further phase separation. 
At the same time, these vortices induce high shear rates in the surrounding fluid, which inhibits unmixing therein.
As a vortex strengthens, the composition in its interior approaches $\varphi\sim\pm1$, while that of the exterior remains largely homogeneous with $|\varphi|\ll1$.

Following the initial nucleation phase, vortices move and deform in the swirling flows induced by their neighbors.
Some are destroyed by strong shear due to larger neighbors; others grow and merge to form larger and more powerful vortices.
In this way, the number of vorticies decreases in time until only two large, counter-rotating vortices remain (Fig. \ref{fig:Fig5}). 
Together, these vortices form a stable, self-propelled vortex doublet of size $R$ that swims about the domain with velocity $U\sim \alpha R$.
The doublet creates a velocity disturbance that decays as $\alpha R^3/r^2$ with distance $r$ from its center.
Consequently, a single vortex doublet can create shear rates of order unity (that necessary for mixing) at distances of $R \alpha^{1/3}$.
This result is consistent with the simulation results shown in Figure \ref{fig:Fig5}, in which a single vortex doublet effectively mixes a region \textit{ca.} four times as large as itself with $\alpha=50$.

\subsection{\label{sec:level35}Results of the microscopic model}

\begin{figure}[!htb]

  \includegraphics{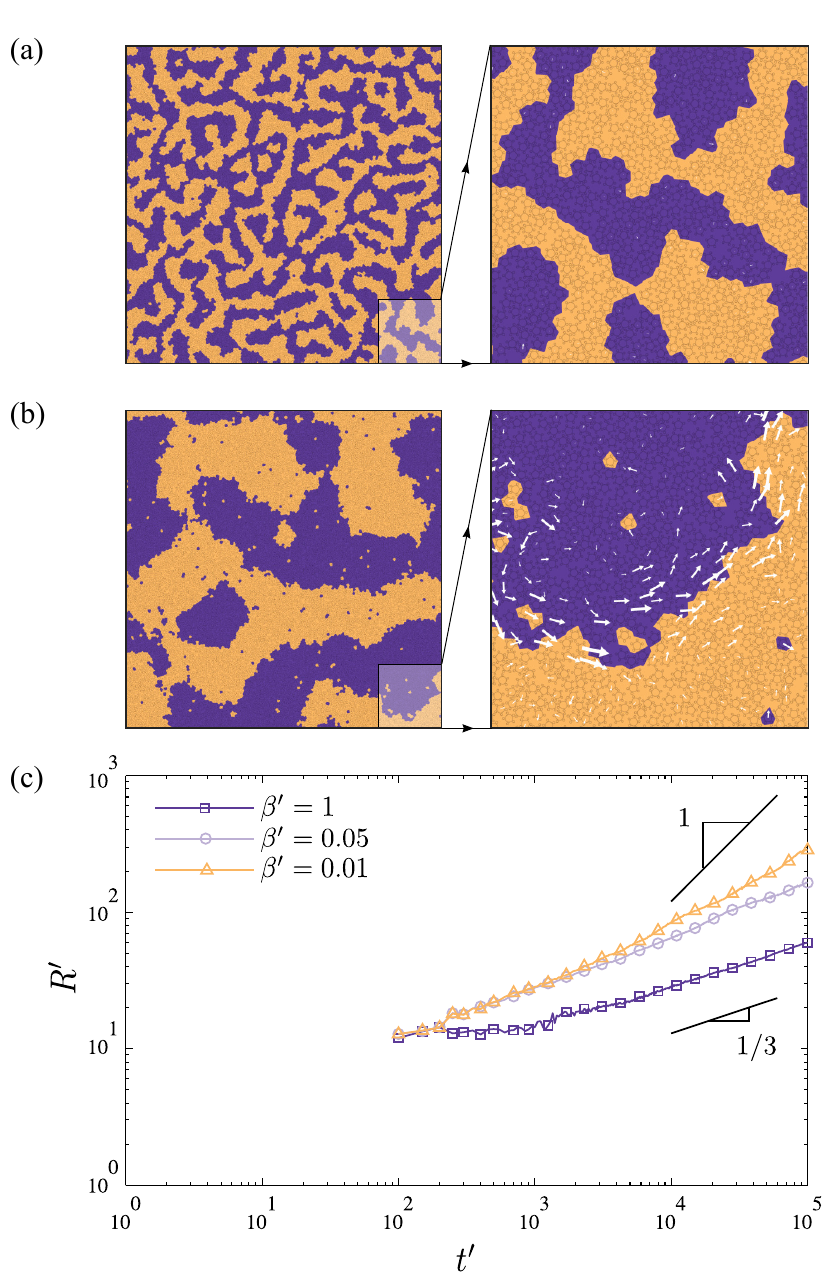}
  \caption{\textbf{Microscopic model.} (a,b) Representative snapshots of a 50:50 mixture of $16,384$ spinners driven to rotate in clockwise (orange) and counterclockwise (purple) directions at time $t' = 10^5$. The visualization of the particles by their voronoi tessellation is shown in the right images along with the fluid flows. The image in (a) show ``passive coarsening'' for parameters $\alpha' = 0.25$ and $\beta' = 1$ corresponding to strong frictional damping.  Image in (b) show ``active coarsening'' for parameters $\alpha' = 0.25$ and $\beta' = 0.01$ corresponding to weak damping. White arrows show the direction of fluid flows. (c) Domain size $R'$ as a function of time for $\alpha' = 0.25$ and different amounts of frictional damping $\beta' = 1$, $0.05$, and $0.01$; this plot is analogous to that in Fig. \ref{fig:Fig3}c. Here, $R'$ is defined as the first zero of the integral of the pair correlation function.}
  \label{fig:Fig6}
\end{figure}

The key insight suggested by the continuum model is that phase separation in actively rotating liquids can drive convective flows that feedback into the system and direct the dynamic evolution of the growing phases.
This result is further supported by microscopic simulations of actively rotating particles (Fig. \ref{fig:Fig6}).
In particular, we studied the collective dynamics of $16,384$ spinners subject to moderate driving torques ($\alpha' = 0.25$) and different levels of frictional damping.
Under these conditions, spinners unmix to form domains of like-rotating particles (Fig. \ref{fig:Fig6}a,b) that grow steadily in time.
Here, the size $R'$ of the growing domains is quantified as the first root of the \textit{integral} of the pair correlation function.
Note that this measure is different from that used in the continuum model and is chosen for its decreased sensitivity to statistical fluctuations.

For relatively strong damping ($\beta' = 1$), the domain size increases as $R' \propto t'^{1/3}$ (Fig. \ref{fig:Fig6}a,c) in agreement with the continuum model (Fig. \ref{fig:Fig2}) and with previous microscopic results \cite{Nguyen2014}.
By contrast, as the damping parameter is decreased, convective flows emerge and accelerate the rate of coarsening (Fig. \ref{fig:Fig6}b).
Consistent with the continuum model, the growth exponent increases from $1/3$ toward $1$ as the damping parameter $\beta'$ is reduced.
The exploration of smaller damping parameters and/or longer simulations times in the microscopic model was found to be computationally prohibitive.
Therefore, it is unclear if the microscopic model will approach a scaling exponent of \textit{ca.} $1$ in the limit as $\beta' \rightarrow 0$ or if it will return to a scaling exponent of $1/3$ in the limit of long times.

To explore the possibility of ``vortex doublets'' in the microscopic model, we increased the driving torque to $\alpha' = 2.5$ under conditions of weak damping ($\beta' = 0.01$).  
Consistent with the continuum model, the increased rotational activity of the particles was sufficient to inhibit the unmixing of the spinners; however, we did not observe the nucleation of localized vortices from homogeneous initial conditions. 
Furthermore, we applied the same driving torque to an initially phase-separated system under conditions of weak damping ($\alpha' = 10$ and $\beta' = 0.01$). 
The active rotation of the particles resulted in their complete mixing, which suggests that the absence of vortex doublets in the microscopic model is not the result of a nucleation barrier.

The discrepancies between the two models under conditions of high torque likely arise from a failure of the continuum model to account for the microscopic effects of active rotation on fluid-fluid phase separation.
In the microscopic model, the driving torque $\alpha'$ sets an energy scale, which must be significantly less than that of attractive interparticle interactions to achieve microscopic phase separation.
In the high-torque simulations, these two energy scales are comparable such that active rotation prohibits any and all phase separation.
In such systems, the strong torques required for the formation of vortex structures lead instead to the complete mixing of the binary fluid.
Additional theoretical work is required to account for the microscopic details of the particle-based simulations (e.g., particle shape) within an accurate hydrodynamic description.
Nevertheless, it remains likely that the vortex structures observed in the hydrodynamic model could also be realized in analogous microscopic models provided the driving force for phase separation is sufficiently strong. 

\section{\label{sec:level4}Conclusions}
To summarize, we presented a phenomenological, continuum model for studying the dynamics of phase separation in binary liquids with active rotation. 
Convective flows induced by the fluid activity result in accelerated coarsening as compared to spinodal decomposition in passive liquids.
The transition from passive to active coarsening is determined primarily by the strength of active rotation and that of frictional interactions with the stationary surroundings.
In addition to active coarsening, the continuum model also predicts the formation of self-propelled vortex doublets under conditions of strong rotation and weak frictional damping.
These dissipative structures emerge spontaneously from the competition between fluid mixing via active rotation and fluid unmixing due to interparticle interactions.
Many of the trends observed in the continuum model such as accelerated coarsening are also reproduced by microscopic kinetic simulations of counter-rotating particle mixtures.
Further work is needed to develop a more rigorous connection between such microscopic models and the continuum hydrodynamics of actively rotating fluids.
We are currently developing experimental models of counter-rotating particle mixtures to explore and elucidate the collective dynamics that emerge from active rotation.
Such nonequilibrium materials have the potential to achieve life-like properties such the abilities to adapt, reconfigure, and repair dynamically in response to environmental stimuli.

\begin{acknowledgments}
This work was supported as part of the Center for Bio-Inspired Energy Science, an Energy Frontier Research Center funded by the U.S. Department of Energy, Office of Science, Basic Energy Sciences under Award DE-SC0000989.
\end{acknowledgments}

\bibliography{References.bib}
\end{document}